\documentclass[reprint,aps,ams,amsmath,prx,twocolumn,superscriptaddress,longbibliography]{revtex4-1}
\usepackage{graphics}
\usepackage{graphicx}
\usepackage{epsfig}
\usepackage{amsmath}
\usepackage{amssymb}
\usepackage{amsfonts}
\usepackage{bm}
\usepackage{color}
\usepackage{bbm}
\usepackage{hyperref}
\usepackage[normalem]{ulem}
\usepackage{comment}
\usepackage{soul}

\begin{document}

\title{Hydrodynamic Coulomb drag in odd electron liquids}

\author{Dmitry Zverevich}
\affiliation{Department of Physics, University of Wisconsin-Madison, Madison, Wisconsin 53706, USA}

\author{Dmitri B. Gutman}
\affiliation{Department of Physics, Bar-Ilan University, Ramat Gan, 52900, Israel}

\author{Alex Levchenko}
\affiliation{Department of Physics, University of Wisconsin-Madison, Madison, Wisconsin 53706, USA}

\date{June 28, 2025}

\begin{abstract}
We consider the problem of Coulomb drag resistance in bilayers of electron liquids with spontaneously broken time-reversal symmetry. In the hydrodynamic regime, the viscosity tensor of such fluids has a nonvanishing odd component. In this scenario, fluctuating viscous stresses drive the propagation of plasmons, whose dispersion relations are modified by nondissipative odd viscous waves. Coulomb coupling of electron density fluctuations induces a drag force exerted by one layer on the other in the presence of a steady flow. This drag force can be expressed through the dynamic structure factor of the electron liquid, which is peaked at frequencies corresponding to plasmon resonances in the bilayer. As a result, the drag resistivity depends on the dissipationless odd viscosity of the fluid. We quantify this effect and present a general theory of hydrodynamic fluctuations applicable to odd electron liquids, both with and without Galilean invariance.      
\end{abstract}

\maketitle

\section{Introduction}

Odd fluids and odd rigid media are characterized by additional linear response coefficients, known as odd viscosity and odd elasticity, which arise in systems where time-reversal symmetry is broken, either spontaneously or due to an external magnetic field. These systems encompass a wide range of phenomena, from quantum liquids, biological and active matter, to astrophysical gases and beyond (see Ref. \cite{OddReview} for a representative review and references therein).

In the context of electron liquids in solids, odd viscosity naturally arises as a property of an electron gas in a magnetic field. It is therefore usually called Hall viscosity. The semiclassical theory of this quantity was developed long ago based on the Boltzmann equation \cite{Steinberg:1958}. It also appears in the context of plasma physics \cite{Kaufman:1960}. Interest in Hall viscosity grew significantly with the realization that, in a gapped system, it is topologically quantized \cite{Avron:1995,Avron:1998}.

Recent progress in fabrication of high-mobility graphene devices have enabled measurements of electron transport in the hydrodynamic regime (for reviews on the topic see Refs. \cite{Lucas:2018,Levchenko:2020,Narozhny:2022,Fritz:2024}). These advances have sparked significant attention and theoretical studies of magnetotransport in finite geometries, including Hall-bar and Corbino devices, with a focus on the manifestations of Hall viscosity in current flow profiles and magnetoresistance \cite{Hoyos:2012,Vignale:2016,Scaffidi:2017,Delacretaz:2017,Polini:2017,Holder:2019}. Measurements of Hall viscosity were reported recently \cite{Berdyugin:2019}. Interestingly, the most recent findings indicate that the electron liquid in graphene may spontaneously break time-reversal and inversion symmetry. This was realized in the discovered quarter-metal state in rhombohedral trilayer graphene \cite{Zhou:2021}. The time-reversal symmetry breaking is caused by valley and spin polarization. 
These results motivate the question about manifestations of intrinsic odd viscosity in transport properties of electron liquids in scenarios with spontaneous time-reversal symmetry breaking. 

In this work, we provide one such example by considering an electronic double-layer system in a Coulomb drag setup \cite{Review:2016}. In this geometry, the flow of electron liquid in one layer (the drive layer) induces a drag force on the other layer (the drag layer) due to interlayer Coulomb coupling. Under open-circuit conditions in the drag layer, the resulting voltage buildup generates a force that compensates the drag force. Consequently, Coulomb drag resistivity is typically defined as the ratio of this induced drag voltage to the current in the drive layer.

In the hydrodynamic regime, the leading contribution to the drag force arises from plasmon resonances in the bilayer \cite{Apostolov:2014}. We find that in odd electron liquids, the dispersion relation of plasmon modes is modified by odd viscosity. At charge neutrality, we recover the viscosity waves with a quadratic dispersion predicted by Avron \cite{Avron:1998}. At finite density, acoustic and optical plasmons retain their usual dispersion in the long-wavelength limit. However, plasmon poles in the dynamical structure factor of the fluid are modified by odd viscosity, therefore the temperature and interlayer separation dependence of the drag resistance become sensitive to odd viscosity. We quantify this effect in the present paper.     

We find it most convenient to work within the framework of stochastic hydrodynamic theory -- specifically, hydrodynamic equations that incorporate random Langevin fluxes, whose correlation function is fixed by the fluctuation-dissipation relations. This formulation of hydrodynamic theory allows for a straightforward evaluation of correlation functions, such as the density-density correlation function that determines the drag force. This approach was originally developed for classical fluids \cite{LL:1957} and later generalized to superfluids \cite{Khalatnikov:1958}. In a fluid dynamics a thorough presentation of this theory can be found in the textbooks \cite{Forster,LL-V9}. 
It can be further generalized to the kinetic regime within the framework of the Boltzmann-Langevin kinetic equation \cite{Kogan:1996}. In the context of the drag problem, this approach was implemented in Ref. \cite{Chen:2015}.

In Sec. \ref{sec:Hydro}, we provide a concise formulation of this theory, incorporating extensions relevant to odd liquids through their stress tensor. We make no specific assumptions about the microscopic origin of spontaneous time-reversal symmetry breaking or the presence of Galilean invariance. As a result, our findings apply to a broad class of electron liquids that lack Galilean invariance. We identify quantitative differences in drag resistivity between liquids with and without Galilean invariance, as discussed in Sec. \ref{sec:Drag}. In Sec. \ref{sec:Summary}, we summarize our main results and provide a broader discussion in the context of existing related works on the subject.

\section{Hydrodynamic fluctuations in odd electron liquids}\label{sec:Hydro}

Hydrodynamic theory describes the propagation of conserved quantities such as particle density $n$, momentum $\bm{p}$, and energy $\varepsilon$ \cite{LL-V6}. In practical applications, it is often more convenient to work with entropy density $s$ rather than energy.

In the context of electron liquids, the hydrodynamic limit applies when the electron mean free path due to electron-electron collisions, $l_{\text{ee}}$, is much shorter than the mean free path associated with electron-impurity $l_{\text{ei}}$ and electron-phonon $l_{\text{ep}}$ scattering. These scattering processes relax momentum and energy of an electron fluid. Achieving this regime is challenging, as it requires extremely high-quality samples.
For an electron bilayer, where two electron liquids are separated by a distance $d$, the hydrodynamic regime also requires $d>l_{\text{ee}}$.

To have a concise presentation we introduce column-vector notations 
\begin{equation}
\vec{x}=\left(\begin{array}{c} n \\ s \end{array}\right),\quad \vec{\bm{J}}=\left(\begin{array}{c} \bm{j}_n \\ \bm{j}_s \end{array}\right), \quad 
\vec{\bm{X}}=\left(\begin{array}{c} -e\bm{\mathcal{E}} \\ \bm{\nabla}T \end{array}\right)
\end{equation}
for particle and entropy densities $\vec{x}$, their respective currents $\vec{\bm{J}}$, and thermodynamically conjugated forces $\vec{\bm{X}}$. 
The latter consists of electromotive force (EMF) and the local temperature gradients. 
The bold faces denote vectors in two-dimensional space, while vector symbol denotes vector columns.   
In these notations, conservation of particle number and entropy are expressed by
the continuity equation
\begin{equation}\label{eq:dxdt}
\partial_t\vec{x}+\bm{\nabla}\cdot\vec{\bm{J}}=0. 
\end{equation}
The constitutive relation for the current densities takes the form 
\begin{equation}\label{eq:J}
\vec{\bm{J}}=\bm{v}\vec{x}-\hat{\Upsilon}\vec{\bm{X}}+\vec{\bm{I}}.
\end{equation}
It is important to emphasize that in Galilean-invariant liquids, the particle current density is uniquely determined by the local hydrodynamic velocity $\bm{v}$, which, in turn, is defined by the momentum density. This dependence is captured by the first term in the equation above. In the absence of Galilean invariance, additional dissipative contributions to the particle current arise, as described by the second term, where
\begin{equation}
\hat{\Upsilon}=\left(\begin{array}{cc}\sigma/e^2 & \gamma/T \\ \gamma/T & \kappa/T \end{array}\right).
\end{equation}
The matrix of kinetic coefficients $\hat{\Upsilon}$ characterizes the dissipative properties of the electron liquid and is described by intrinsic conductivity $\sigma$, thermal conductivity $\kappa$, and the thermoelectric coefficient $\gamma$. For example, for a graphene monolayer intrinsic conductivity was calculated in Refs. \cite{Fritz:2008,Kashuba:2008}. Particle-hole symmetry requires the intrinsic thermoelecric coefficient to vanish at charge neutrality $n\to0$. It can be estimated to scale as $\gamma/T\propto n/s\ll1$. The last term in Eq. \eqref{eq:J} describes the random flux of currents. These fluxes are thermally driven and describe spatial and temporal fluctuations. Through the Onsager principle and the fluctuation-dissipation theorem (FDT), the correlation function of these fluxes is determined by the matrix of dissipative coefficients 
\begin{equation}\label{eq:I}
\langle \vec{\bm{I}}(\bm{r},t)\otimes\vec{\bm{I}}^{\mathbb{T}}(\bm{r}',t')\rangle=2T\hat{\Upsilon}\delta(\bm{r}-\bm{r}')\delta(t-t'). 
\end{equation}
Here $\langle\ldots\rangle$ denotes averaging over the thermal fluctuations, 
the notation $\otimes$ is used to denote the direct product of two vectors, whereas superscript $\mathbb{T}$ denotes vector transposition.

The evolution equation for the momentum density can be written in the form of Newton's second law
\begin{equation}\label{eq:p}
\partial_t\bm{p}=-\bm{\nabla}\cdot\hat{\Pi}-en\bm{\nabla}\Phi
\end{equation}
In this equation, the self-consistent electric potential $\Phi$ is related to the electron density through the Poisson equation. Its presence reflects the flow of momentum in the electron fluid due to long-range Coulomb interactions between electrons. For a bilayer system, $\Phi$ includes the full potential generated by density fluctuations in both layers. The local part of the momentum flux is described by the tensor \cite{LL-V6,LL-V9}
\begin{equation}
\hat{\Pi}\equiv\Pi_{ij}=P\delta_{ij}+\varrho v_iv_j-\Sigma_{ij}, 
\end{equation}
where $\varrho=mn$ is the mass density of the fluid, which includes the local hydrodynamic pressure $P$ and viscous stress tensor
\begin{equation}
\Sigma_{ij}=\Sigma^e_{ij}+\Sigma^o_{ij}+\Xi_{ij}.
\end{equation}
It can be split into three parts. The first one (even term) takes the usual form
\begin{equation}
\Sigma^e_{ij}=2\eta V_{ij}+(\zeta-\eta)\delta_{ij}\partial_kv_k, 
\end{equation} 
which is described by the dissipative shear $\eta$ and bulk $\zeta$ viscosities, with $V_{ij}=(\partial_iv_j+\partial_jv_i)/2$ and 
$\delta_{ij}$ being Kroenecker delta symbol. Here $i,j$ are Cartesian indices, and we used shorthand notation for the spatial derivative $\partial_i=\partial/\partial x_i$. The summation over the repeated indices is implicit. The second odd term is given by  
\begin{equation}
\Sigma^o_{ij}=\eta_o(\epsilon_{ik}V_{jk}+\epsilon_{jk}V_{ik}), 
\end{equation} 
where $\epsilon_{ij}$ is the 2D antisymmetric Levi-Civita tensor, and $\eta_o$ is the dissipationless odd viscosity. In these expressions we assume isotropic fluid. The last third term describes thermally-driven fluctuations of viscous stresses $\Xi_{ij}$ whose correlation function is given by 
\begin{align}\label{eq:Xi}
\langle\Xi_{ik}(\bm{r},t)\Xi_{lm}(\bm{r}',t')\rangle=2T\delta(\bm{r}-\bm{r}')\delta(t-t')\nonumber \\ 
\times [\eta(\delta_{il}\delta_{km}+\delta_{im}\delta_{kl})+(\eta-\zeta)\delta_{ik}\delta_{lm}].
\end{align}
Following the approach of Ref. \cite{LL:1957}, it can be readily verified that odd viscosity does not enter the correlation function in Eq. \eqref{eq:Xi}. This is yet another manifestation of its nondissipative nature, as it drops out of the fluctuation-dissipation theorem. Using the same line of reasoning, one can also confirm that odd viscosity cannot lead to an increase in the liquid’s temperature, nor does it affect the evolution of entropy due to heat conduction, regardless of whether the liquid is compressible or incompressible. Consequently, energy dissipation is governed in the usual manner -- solely through the even part of the stress tensor \cite{LL-V6}. Previous studies have shown additionally that, for fluid flows in confined geometries, the velocity field -- depending on the boundary conditions -- remains independent of odd viscosity. Similarly, the force acting on a closed contour is also unaffected by odd viscosity \cite
{Ganeshan:2017}. For these reasons, the manifestations of odd viscosity in electronic transport are rather subtle.

\section{Coulomb drag resistance}\label{sec:Drag}

In this section, we apply the formalism of hydrodynamic fluctuations to evaluate the drag force and drag resistance in a bilayer of odd electron liquids. 

The physical picture of Coulomb drag effect in the hydrodynamic regime can be understood as follows. Thermally induced fluctuations of viscous stresses and intrinsic currents drive electron density fluctuations in the fluid. These fluctuations propagate as plasmons. In the presence of a steady current $\bm{j}$ in the active layer, density fluctuations are advected by the flow. Coulomb coupling of these fluctuations between the layers results in a drag force $\bm{F}_{\text{D}}$. This force can be determined by relating the potential to density fluctuations using the Poisson equation. The result is \cite{Apostolov:2014}
\begin{equation}\label{eq:F}
\bm{F}_{\text{D}}=\int\frac{d^2qd\omega}{(2\pi)^3}(-i\bm{q})\left(\frac{2\pi e^2}{\varkappa q}\right)e^{-qd}D(\bm{q},\omega),
\end{equation}     
where $\varkappa$ is the dielectric constant of the host material surrounding the
electron layers and 
\begin{equation}
D(\bm{q},\omega)=\langle\delta n_1(\bm{q},\omega)\delta n_2(-\bm{q},-\omega)\rangle
\end{equation}
is the interlayer density-density correlation function (dynamical structure factor). Here $\delta n_{1,2}(\bm{q},\omega)$ are the Fourier components of the density
fluctuations in both layers. Provided that $D(\bm{q},\omega)$ is known to the liner order in $\bm{v}$ the drag resistivity is given by 
\begin{equation}\label{eq:rho}
\rho_{\text{D}}=\frac{\bm{v}\cdot\bm{F}_{\text{D}}}{(env)^2}.
\end{equation}

The solution strategy to determine $\rho_{\text{D}}$ can be summarized as follows. We linearize Eqs. \eqref{eq:J} and \eqref{eq:p} in fluctuations $\delta n_{1,2}$ and $\delta\bm{v}_{1,2}$. 
Each equation should be replicated to describe fluctuations in both layers labeled by $(\delta n_1,\delta\bm{v}_1)$ for the drive layer, and $(\delta n_2,\delta\bm{v}_2)$ for the drag layer. In the drive layer we also assume a steady flow with velocity $\bm{v}$ with fluctuations $\delta\bm{v}_1$ occurring on top of it. For simplicity we take that the average electron density is the same in each layer $n_{1,2}=n$. We further assume identical screening properties of both layers.  
We solve these equations to determine fluctuations in equilibrium first and then iterate this solution once to determine nonequilibrium parts of density fluctuations to the linear order in $\bm{v}$. 
Thermal averages leading to the dynamic structure factor can be calculated with the help of the correlations functions defined in Eqs. \eqref{eq:I} and \eqref{eq:Xi}.  
We implement this strategy by breaking down solutions separately for Galilean-invariant systems and those without Galilean invariance.

\subsection{Drag in Galilean-invariant systems}

The starting point of our derivation is the linearized continuity equation \eqref{eq:J}. For Galilean invariant systems, we set $\sigma\to0$ and $\gamma\to0$, and thus obtain a standard expression for the Fourier components of fluctuating quantities $\delta n,\delta\bm{v}\propto e^{-i\omega t+i\bm{qr}}$. For the drive layer we find 
\begin{equation}\label{eq:dndv}
-\omega\delta n_1+(\bm{q}\cdot\bm{v})\delta n_1+n(\bm{q}\cdot\delta\bm{v}_1)=0.
\end{equation} 
The linearized continuity equation for the momentum density Eq. \eqref{eq:p} can be conveniently written in projection components of velocity field fluctuations parallel and tangential to directions of $\bm{q}$, namely 
\begin{equation}\label{eq:v-projections}
\delta v^{\parallel}=\frac{(\bm{q}\cdot\delta\bm{v})}{q},\quad \delta v^\perp=\frac{(\bm{q}\cdot[\delta\bm{v}\times\hat{\bm{z}}])}{q},
\end{equation}
where $\hat{\bm{z}}$ is the unit vector along the $z$-direction perpendicular to the plane of a 2D electron system, which can be associated a spatial direction
of the breaking of time-reversal symmetry. For these components we find 
\begin{subequations}\label{eq:linearized-NS}
\begin{align}
&\varrho[-i\omega+i(\bm{v}\cdot\bm{q})]\delta v^\parallel_1=\nonumber\\ 
&-iqne\delta\Phi_1-\eta q^2\delta v^\parallel_1-\eta_oq^2\delta v^\perp_1+\frac{i}{q}(\bm{q}\cdot\hat{\Xi}_1\bm{q}), \\ 
&\varrho[-i\omega+i(\bm{v}\cdot\bm{q})]\delta v^\perp_1=
\nonumber\\ 
&-\eta q^2\delta v^\perp_1+\eta_oq^2\delta v^\parallel_1+\frac{i}{q}(\bm{q}\cdot[\hat{\Xi}_1\bm{q}\times\hat{\bm{z}}]), 
\end{align}
\end{subequations}
where we used the shorthand notation $(\bm{q}\cdot\hat{\Xi}_1\bm{q})\equiv q_iq_j(\Xi_{ij})_1$ and analogously for the vector product quantities. 
The coupling between the layers enters through the Coulomb potential
\begin{equation}\label{eq:Phi}
e\delta\Phi_1=\frac{2\pi e^2}{\varkappa q}(\delta n_1+e^{-qd}\delta n_2).
\end{equation} 
For the drag layer these expressions are the same, one only needs to flip indices $1\leftrightarrow2$ and set $\bm{v}\to0$. 

A few comments are in order regarding the form of Eq.~\eqref{eq:linearized-NS}, which we derived while making two additional simplifying approximations. First, we neglected terms involving bulk viscosity. These terms appear in the equation for $\delta v^\parallel$ in the combination $(\eta+\zeta)q^2$ but do not enter the equation for $\delta v^\perp$. This is a reasonable approximation, as bulk viscosity is typically much smaller than shear viscosity. For instance, it is known that in systems with linear and quadratic dispersion, bulk viscosity vanishes \cite{LL-V10}. Second, we omitted terms proportional to the gradients of pressure fluctuations, $\bm{\nabla}\delta P$, which arise from both density and entropy fluctuations. This is justified because the long-range nature of the Coulomb potential in Eq.~\eqref{eq:Phi} dominates fluctuations at long wave length $q\to0$. As a result, the equation governing entropy fluctuations decouples. This neglects contributions to the drag arising from changes in fluid density due to heat fluxes. However, previous estimates suggest that these contributions are smaller than those driven by viscous stresses across the entire temperature range of the collision-dominated regime \cite{Chen:2015}.

As the next step, we use the continuity equation to eliminate velocity fluctuations, $\delta v^\parallel$ and $\delta v^\perp$. This allows us to derive a coupled set of equations for density fluctuations in both layers. We then introduce symmetric and antisymmetric combinations, $\delta n_\pm=\delta n_1\pm\delta n_2$, along with the corresponding Langevin fluxes, $\hat{\Xi}_\pm=\hat{\Xi}_1\pm\hat{\Xi}_2$. This transformation clarifies the physical picture by capturing propagating in-phase and out-of-phase density modes. Thus, by combining Eqs.~\eqref{eq:dndv} and \eqref{eq:linearized-NS} and implementing the steps outlined above, we obtain the density response in a compact form:
\begin{align}\label{eq:delta-n}
&\delta n_\pm=\omega_o\frac{q^2\Xi^\perp_\pm}{m\mathcal{P}_\pm}-(\omega_\eta-i\omega)\frac{q^2\Xi^\parallel_\pm}{m\mathcal{P}_\pm}\nonumber \\ 
&+\frac{i(\bm{v}\cdot\bm{q})}{2\mathcal{P}_\pm}\left[\Gamma_+\delta n_++\Gamma_-\delta n_--\frac{q^2(\Xi^\parallel_++\Xi^\perp_-)}{m}\right].
\end{align}
Here the projectors of Langevin fluxes are defined in conjunction with Eq. \eqref{eq:v-projections}, namely 
\begin{equation}
\Xi^\parallel_\pm=\frac{\bm{q}\cdot(\hat{\Xi}_\pm\bm{q})}{q^2},\quad \Xi^\perp_\pm=\frac{\bm{q}\cdot[\hat{\Xi}_\pm\bm{q}\times\hat{\bm{z}}]}{q^2}.
\end{equation}
We have also introduced the polarization function whose form is given by 
\begin{equation}\label{eq:P}
\mathcal{P}_\pm=i\omega(\omega_\eta-i\omega)^2-(\omega_\eta-i\omega)\omega^2_\pm+i\omega\omega^2_o,
\end{equation}  
and dynamic vertex functions, 
\begin{equation}\label{eq:Gamma}
\Gamma_\pm=(\omega_\eta-i\omega)^2-2i\omega_\eta+\omega^2_\pm+\omega^2_o,
\end{equation}
that couple equilibrium fluctuating density modes in the presence of the hydrodynamic flow. 

The relevant energy scales in this problem include the following quantities 
\begin{align}
\omega_\pm=\omega_{\text{p}}\sqrt{1\pm e^{-qd}},\quad \omega_{\text{p}}=\sqrt{\frac{2\pi ne^2q}{m\varkappa}},\nonumber \\ 
 \omega_\eta=\nu q^2,\quad \omega_o=\nu_oq^2. 
\end{align}  
Here $\omega_\text{p}$ is the dispersion of a 2D plasmon whereas $\omega_\pm$ are symmetric/antisymmetric plasmon frequencies in a bilayer. In addition, $\nu=\eta/\varrho$ is the dissipative kinematic viscosity which defines diffusive-like spreading of charge fluctuations characterized by the scale $\omega_\eta$, and $\nu_o=\eta_o/\varrho$ is the dissipationless kinematic viscosity defining $\omega_o$.  

To understand the different roles of these energies, it is useful to examine the zeros of the polarization function 
$\mathcal{P}_\pm$, which defines the collective modes of the system \cite{Forster}. For this purpose, let us momentarily switch off dissipation, setting 
$\omega_\eta\to0$. In this limit, Eq.~\eqref{eq:P} simplifies to $\mathcal{P}_\pm=-i\omega(\omega^2-\omega^2_\pm-\omega^2_o)$. 
The collective mode is thus defined by the relation 
\begin{equation}\label{eq:mode}
\omega=\sqrt{\omega^2_\pm+\omega^2_o}. 
\end{equation}
At charge neutrality, $n\to0$, this coincides with the odd viscosity wave, which exhibits quadratic dispersion 
$\nu_oq^2$, as predicted by Avron \cite{Avron:1998}. Conversely, in the absence of odd viscosity ($\eta_o\to0$), we recover the standard plasmon dispersion. The acoustic mode corresponds to out-of-phase oscillations and therefore exhibits a linear dispersion $\omega\propto q$. The optical plasmon, on the other hand, corresponds to in-phase electron density oscillations, and in the long-wavelength limit, its dispersion relation coincides with that of a single-layer plasmon, following a square-root dependence $\omega\propto \sqrt{q}$. 
Next, restoring dissipation while keeping odd viscosity at zero, we obtain $\mathcal{P}_\pm=(\omega_\eta-i\omega)[\omega^2-\omega^2_\pm+i\omega\omega_\eta]$.
The first term, $\omega_\eta-i\omega$, corresponds to a viscous diffusion pole. The second term is the familiar plasmon pole, now broadened by viscous charge-spreading dissipation. This confirms the key result that the lifetime of a plasmon in the hydrodynamic regime of a Galilean-invariant fluid is governed by viscosity, with $\Im\omega=\nu q^2/2$ \cite{Hruska:2002}.
With all terms accounted for, we see that odd viscosity modifies the plasmon dispersion, while viscous terms determine the lifetime of the collective modes.

We proceed to solve Eq. \eqref{eq:delta-n} in perturbation theory over $\bm{v}$ to the leading linear order. The zeroth order solution, which we denote as $\delta n^{(0)}_\pm$, is obtained at $\bm{v}\to0$. It corresponds to the equilibrium density fluctuations driven by thermal fluctuations of viscous stresses. The first order corrections, which we denote as $\delta n^{(1)}_\pm$, corresponds to the nonequilibrium     
electron density fluctuations advected by the flow. We thus find
\begin{subequations}
\begin{align}
&\delta n_\pm=\delta n^{(0)}_\pm+\delta n^{(1)}_\pm,\\ 
&\delta n^{(0)}_\pm=\omega_o\frac{q^2\Xi^\perp_\pm}{m\mathcal{P}_\pm}-(\omega_\eta-i\omega)\frac{q^2\Xi^\parallel_\pm}{m\mathcal{P}_\pm},\\ 
&\delta n^{(1)}_\pm=\frac{i(\bm{v}\cdot\bm{q})}{2\mathcal{P}_\pm}\left[\Gamma_+\delta n^{(0)}_++\Gamma_-\delta n^{(0)}_--\frac{q^2(\Xi^\parallel_++\Xi^\perp_-)}{m}\right].
\end{align}
\end{subequations}
These expressions enable the calculation of the dynamical structure factor. Using the thermal averages 
\begin{equation}
\langle\Xi^a_i\Xi^b_j\rangle=4T\eta\delta_{ij}\delta_{ab}, \quad (i,j)=\pm, \quad (a,b)=\parallel,\perp
\end{equation}
that follow from Eq. \eqref{eq:Xi} upon a Fourier transform and with the neglect of the bulk viscosity (to be consistent with the previously made approximation), and after tedious, but otherwise straightforward algebra we find 
\begin{equation}
D(\bm{q},\omega)=\frac{i(\bm{v}\cdot\bm{q})nTq^2}{m|\mathcal{P}_+|^2|\mathcal{P}_-|^2}\omega^2_\eta(\omega^2_+-\omega^2_-)(\omega^2+\omega^2_o+\omega^2_\eta)^2.
\end{equation}
At this stage, we combine Eqs. \eqref{eq:F} and \eqref{eq:rho} and the above expression for the structure factor $D(\bm{q},\omega)$ to find the drag resistivity in the form 
\begin{align}
\rho_{\text{D}}=&\frac{1}{2e^2n^2}\int\frac{d\omega d^2q}{(2\pi)^3}q^2\left(\frac{2\pi e^2}{\varkappa q}\right)e^{-qd}\nonumber\\
&\times\frac{nTq^2}{m|\mathcal{P}_+|^2|\mathcal{P}_-|^2}\omega^2_\eta(\omega^2_+-\omega^2_-)(\omega^2+\omega^2_o+\omega^2_\eta)^2.
\end{align}
Even though we cannot obtain a closed analytical expression for the remaining frequency and momentum integrals, we can still express the final result in a compact form. Specifically, by scaling all energies in units of the plasma frequency evaluated at the wave number corresponding to the interlayer separation, and all wave numbers in the same unit, the final result can be rewritten as
\begin{equation}\label{eq:drag-GI}
\rho_{\text{D}}=\frac{1}{2e^2}\left(\frac{\varkappa v_{\text{F}}}{e^2}\right)\frac{\eta/n}{(k_{\text{F}}d)^5}\frac{T}{E_{\text{F}}}f(\beta,\varsigma). 
\end{equation}
Here $k_{\text{F}}$ and $E_{\text{F}}$ are the Fermi momentum and energy, respectively, with the limit $k_{\text{F}}d>1$.  
The two-parameter function is defined by the dimensionless double-integral 
\begin{subequations}\label{eq:f}
\begin{equation}
f(\beta,\varsigma)=\int\limits^{\infty}_0dx\int\limits^{+\infty}_{-\infty}dy
\frac{\beta x^9e^{-2x}(y^2+\beta^2x^4(1+\varsigma^2))^2}{|\pi_+(x,y)|^2|\pi_-(x,y)|^2},
\end{equation}
where 
\begin{equation}
\pi_\pm=iy(\beta x^2-iy)^2-(\beta x^2-iy)x(1\pm e^{-x})+iy\beta^2\varsigma^2x^4
\end{equation}
and dimensionless parameters are 
\begin{equation}
\beta=\frac{\nu}{\varpi d^2},\quad \varsigma=\frac{\eta_o}{\eta}. 
\end{equation} 
\end{subequations}
Here the characteristic energy scale of a plasmon is $\varpi=\sqrt{2\pi ne^2/md\varkappa}$. 

%%%%%%%%%%%%%%%%%%%%%%%%%%%%%%
%%%%%%%%%%%%%%%%%%%%%%%%%%%%%%
\begin{figure}[t!]
\includegraphics[width=\linewidth]{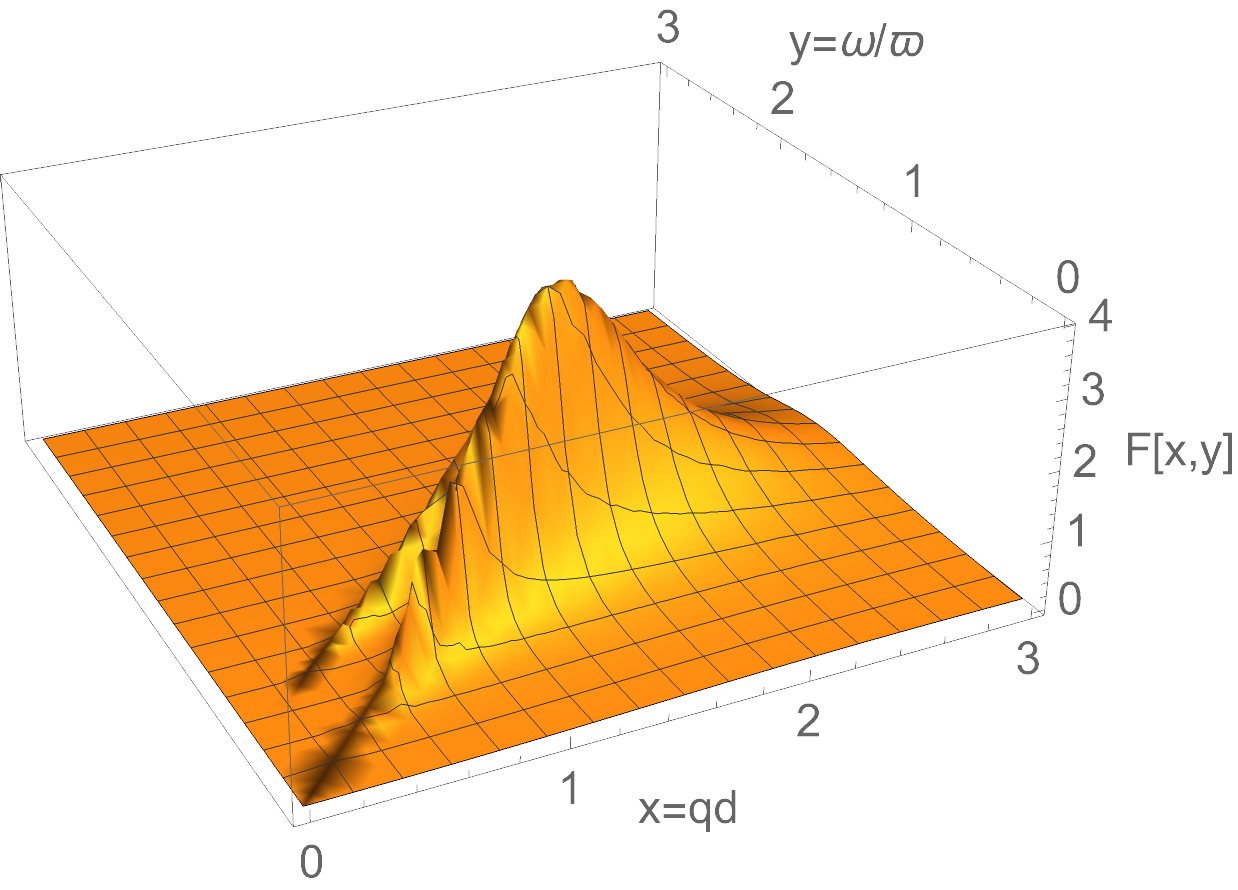}
\caption{Plot of the integrand of Eq. \eqref{eq:f} in the rescaled units $x=qd$ and $y=\omega/\varpi$ representing the spectral weight of a function that defines drag resistance $\rho_{\text{D}}$. On the plot we took $\beta=0.15$ and $\varsigma=0.25$. This pot is qualitatively similar for any values of these parameters  in the range $(\beta,\varsigma)<1$. }\label{fig:3D}
\end{figure}
%%%%%%%%%%%%%%%%%%%%%%%%%%%%%%
%%%%%%%%%%%%%%%%%%%%%%%%%%%%%%

\begin{figure}[t!]
\includegraphics[width=\linewidth]{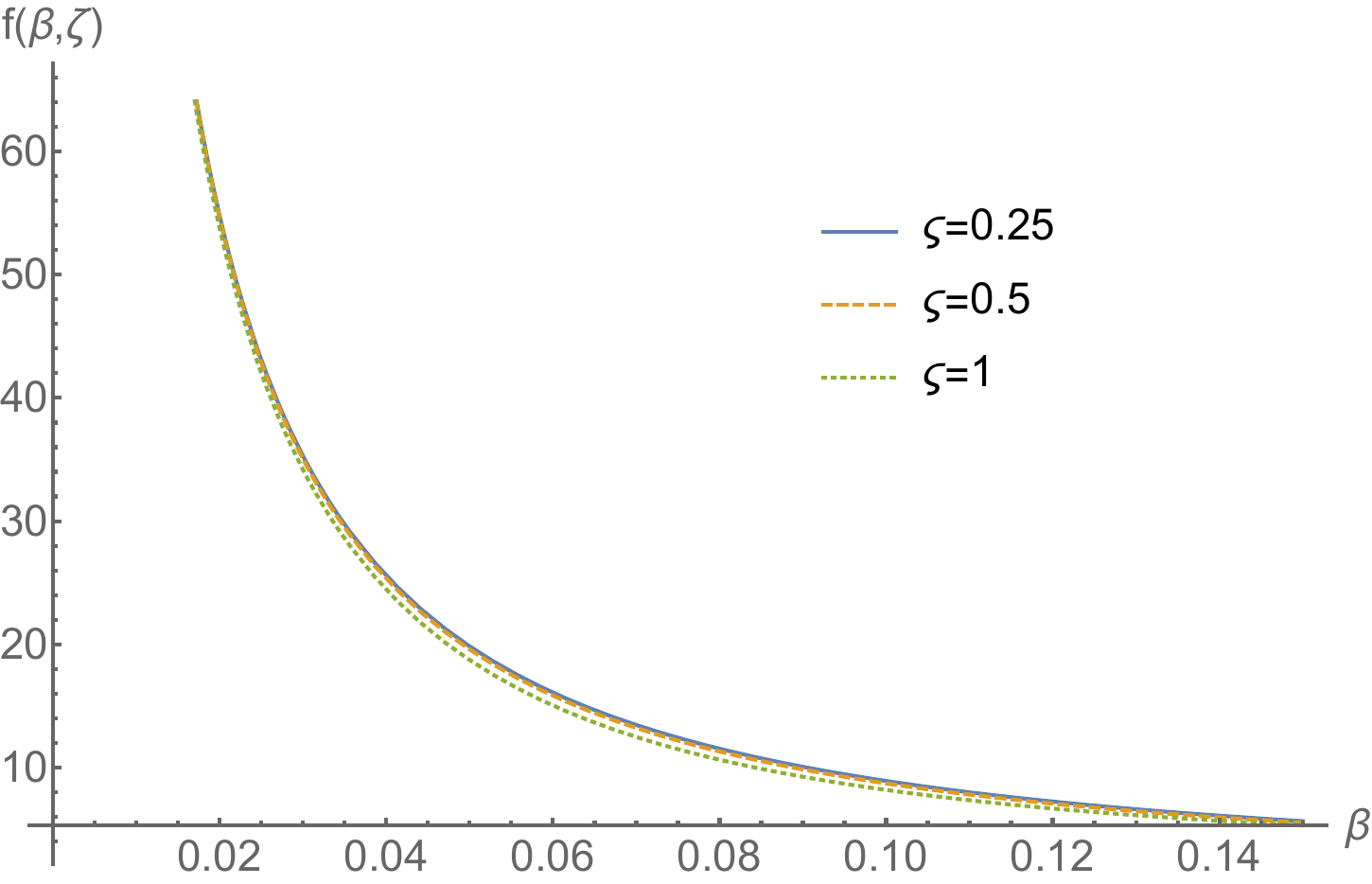}
\caption{Plot of the dimensionless function $f$ from Eq. \eqref{eq:f} versus $\beta$ for several representative values of $\varsigma$ shown on the plot legends.}\label{fig:f}
\end{figure}

It is useful to plot the integrand of the $f$-function in these units, as shown in Fig. \ref{fig:3D}. This function consists of a product of the phase space factor, the strength of the Langevin fluxes, the Coulomb potential, and the dynamical structure factor. In reduced units, the spectral weight is maximal at $x\sim y\sim1$, revealing emerging ridges that correspond to the dispersions of the plasmon modes in a bilayer, $\sqrt{x(1\pm e^{-x})}$, which merge together for $x>1$ when splitting between them becomes exponentially small. The function can be evaluated numerically, and the resulting plot is shown in Fig. \ref{fig:f}.

%%%%%%%%%%%%%%%%%%%%%%%%%%%%%%
%%%%%%%%%%%%%%%%%%%%%%%%%%%%%%
%%%%%%%%%%%%%%%%%%%%%%%%%%%%%%
%%%%%%%%%%%%%%%%%%%%%%%%%%%%%%

\subsection{Drag without Galilean invariance}

\begin{figure}[t!]
\includegraphics[width=\linewidth]{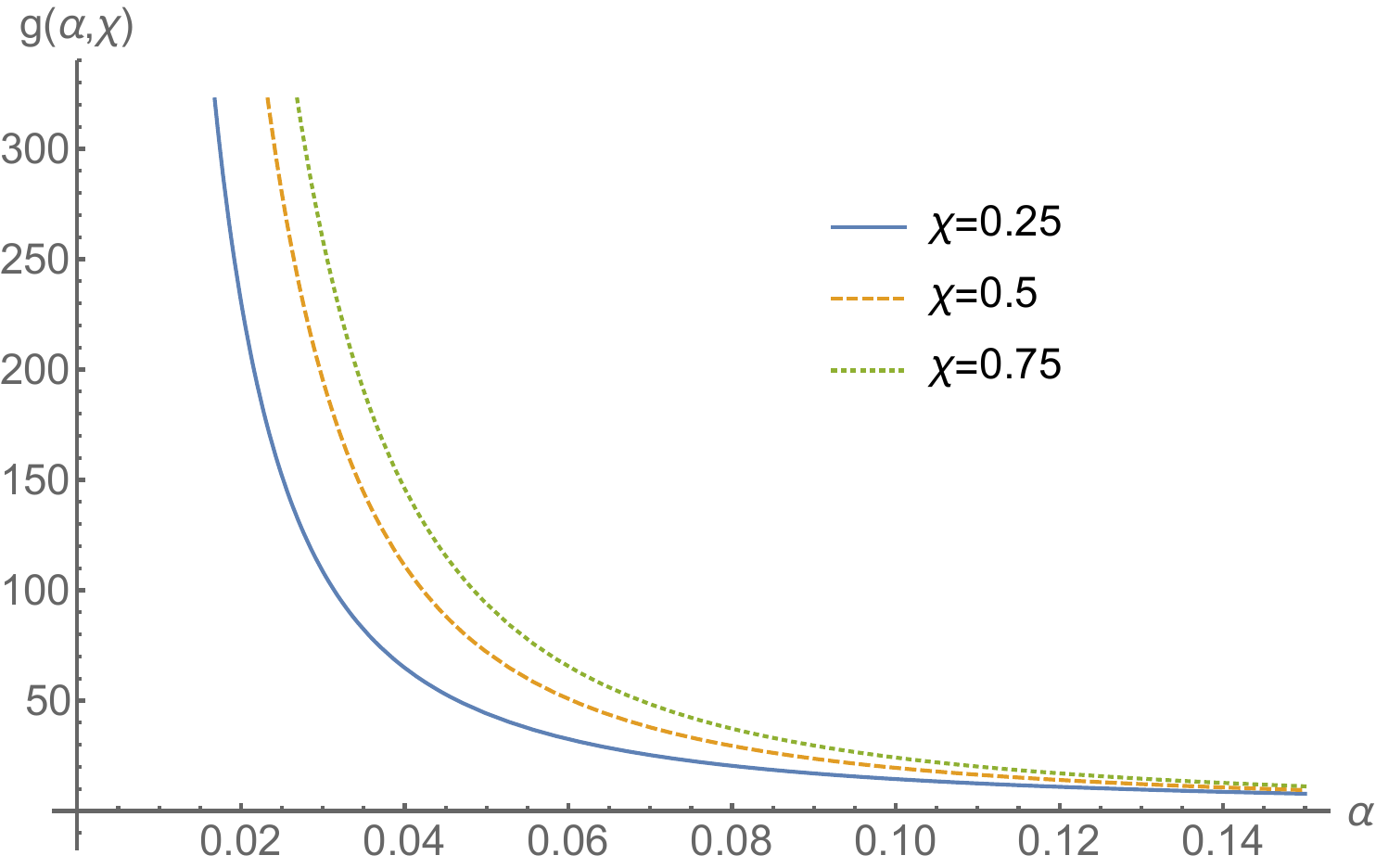}
\caption{Plot of the dimensionless function $g$ from Eq. \eqref{eq:g} versus $\alpha$ for several representative values of $\chi\propto\eta_o$ shown on the plot legends.}\label{fig:g}
\end{figure}

In electron systems without Galilean invariance there is an additional source of density fluctuations, which is driven by the processes related to intrinsic conductivity \cite{Patel:2017}. Incorporating these terms modifies the continuity equation for the particle density. We find from Eq. \eqref{eq:dxdt} and \eqref{eq:J} in Fourier components for the drive layer
\begin{align}
-i\omega\delta n_1+i(\bm{q}\cdot\bm{v})\delta n_1&+i(\bm{q}\cdot\delta\bm{v}_1)n\nonumber \\ &+\frac{\sigma}{e^2}q^2e\delta\Phi_1+i(\bm{q}\cdot\bm{I}_1)=0, 
\end{align}
and similarly for the drag layer. This equation needs to be combined with the Navier-Stokes equations \eqref{eq:linearized-NS} and solved to find resulting density fluctuations. Analyzing this problem we find that the main effect of intrinsic conductivity is to modify the dampening of plasmons. We find that instead of viscous effect it will be dominated by the Maxwell mechanisms of charge relaxation.  The corresponding attenuation coefficients for both plasmons branches are given by 
\begin{equation}
\gamma_\pm=\gamma_q(1\pm e^{-qd}),\quad \gamma_q=\frac{2\pi\sigma q}{\varkappa}, 
\end{equation}
which is much stronger than the decay due to viscosity that scales as $q^2$ in the long wave length limit. This observation motivates the sensible approximation of neglecting dissipative viscous terms from   
Eq. \eqref{eq:linearized-NS} in this case. In the limit $\eta\to0$ the resulting algebraic problem of finding the density fluctuations simplifies greatly. In the symmetrized basis it reduces to 
\begin{equation}
\delta n_\pm=-i(\omega^2-\omega^2_o)\frac{(\bm{q}\cdot\bm{I}_\pm)}{\mathcal{P}_\pm}+i\frac{(\bm{q}\cdot\bm{v})}{2}\left[\Gamma_+\delta n_++\Gamma_-\delta n_-\right]
\end{equation}
which replaces Eq. \eqref{eq:delta-n} from the previous section. The polarization functions and the vertex functions modify accordingly to 
\begin{subequations}
\begin{align}
&\mathcal{P}_\pm=-i\omega^3+i\omega(\omega^2_\pm+\omega^2_o)+\omega^2\gamma_\pm-\gamma_\pm\omega^2_o, \\ 
&\Gamma_\pm=-\omega^2+\omega^2_\pm+\omega^2_o.
\end{align}
\end{subequations}
The thermal averages of the fluctuating intrinsic currents is given by  
\begin{equation}
\langle(\bm{q}\cdot\bm{I}_i)(\bm{q}\cdot\bm{I}_j)\rangle=4Tq^2\frac{\sigma}{e^2}\delta_{ij},\quad (i,j)=\pm
\end{equation}
To the linear order in $\bm{v}$ the solution for equilibrium and nonequilibrium density fluctuations is given by
\begin{subequations}
\begin{align}
&\delta n^{(0)}_\pm=-i(\omega^2-\omega^2_o)\frac{(\bm{q}\cdot\bm{I}_\pm)}{\mathcal{P}_\pm}, \\ 
&\delta n^{(1)}_\pm=i\frac{(\bm{q}\cdot\bm{v})}{2}\left[\Gamma_+\delta n^{(0)}_++\Gamma_-\delta n^{(0)}_-\right],
\end{align}
\end{subequations}
and the corresponding dynamic structure factor evaluates to 
\begin{equation}
D(\bm{q},\omega)=i(\bm{q}\cdot\bm{v})T\frac{\sigma}{e^2}q^2(\omega^2-\omega^2_o)^2\frac{\Gamma_+\Re\mathcal{P}_--\Gamma_-\Re\mathcal{P}_+}{|\mathcal{P_+}|^2|\mathcal{P}_-|^2}.
\end{equation}
Finally, the resulting expression for the drag resistivity can be found in the form 
\begin{equation}
\rho_{\text{D}}\!=\!\frac{T\sigma}{2e^4n^2}\!
\int\!\frac{d\omega d^2q}{(2\pi)^3}\!\left(\frac{2\pi e^2}{\varkappa q}\right)\!e^{-qd}\frac{q^2(\omega^2-\omega^2_o)^4(\gamma_+-\gamma_-)}{|\mathcal{P}_+|^2|\mathcal{P}_-|^2}.
\end{equation}
In the dimensionless variables it can be presented as follows 
\begin{equation}\label{eq:drag-NGI}
\rho_{\text{D}}=\frac{\sigma}{4\pi^2e^4}\left(\frac{1}{nd^2}\right)^2\frac{T}{E_{\text{F}}}g(\alpha,\chi).
\end{equation} 
The function $g$ is defined by the following double integral  
\begin{equation}\label{eq:g}
g(\alpha,\chi)=\int\limits^{\infty}_0dx\int\limits^{+\infty}_{-\infty}dy\frac{\alpha x^5e^{-2x}(y^2-\chi^2x^4)^4}{|\pi_+(x,y)|^2|\pi_-(x,y)|^2}
\end{equation}
where 
\begin{equation}
\pi_\pm=\alpha x(1\pm e^{-x})(y^2-\chi^2x^4)-iy(y^2-\chi^2x^4-x(1\pm e^{-x}))
\end{equation}
and parameter are 
\begin{equation}
\alpha=\frac{2\pi\sigma}{\varkappa\varpi d},\quad \chi=\frac{\nu_o}{d^2\varpi}.
\end{equation}
We have evaluated this function numerically and plotted in Fig. \ref{fig:g}. It shows a much more sensitive dependence on the odd viscosity than the previous example. It is also almost an order of magnitude larger numerically for a similar choice of dimensionless parameters. 

To motivate the choice of numerical parameters, it is helpful to provide relevant estimates.
For sufficiently high electron density, $k_{\text{F}}d>1$, the characteristic energy of the plasmon can be estimated as $\varpi\sim\sqrt{r_s}E_{\text{F}}/\sqrt{k_{\text{F}}d}$, where $r_s=e^2/v_{\text{F}}\varkappa$ is the electron gas parameter.  This gives an estimate for $\alpha\sim\frac{\sigma}{e^2}\sqrt{r_s/(k_{\text{F}}d)}$ which is typically smaller than unity for a weakly correlated regime where $r_s\sim1$ and $\sigma\sim e^2$. It is harder to estimate $\chi$ since odd viscosity has not been microscopically derived for electron liquids. 
However, based on the analogy between the disorder-induced skew scattering \cite{Sinitsyn:2007,Konig:2021} and interaction-induced skew scattering processes that are key for odd viscosity \cite{Golub:2022,Messica:2024}, we can estimate $\eta_o/\eta\sim 1/(E_{\text{F}}\tau_{\text{ee}})$. Provided that ordinary kinematic viscosity is $\nu\sim v^2_{\text{F}}\tau_{\text{ee}}$ one finds for $\chi\sim1/(k_{\text{F}}d)^{3/2}\lesssim1$ for $r_s\sim1$. 

%%%%%%%%%%%%%%%%%%%%%%%%%%%%%%
%%%%%%%%%%%%%%%%%%%%%%%%%%%%%%
%%%%%%%%%%%%%%%%%%%%%%%%%%%%%%
%%%%%%%%%%%%%%%%%%%%%%%%%%%%%%

\section{Summary and discussion}\label{sec:Summary}

In this work, we analyzed the effect of odd viscosity on Coulomb drag resistivity in electron bilayers. A conceptually similar problem was previously studied in Ref. \cite{Apostolov:2019}, where the effect of Hall viscosity on magnetodrag was considered. Additional complication exists due to the drag and drag-Hall viscosities, which arise from a change of the stress tensor due to the interlayer Coulomb interactions \cite{Schmidt:2023}. These contributions can strongly affect drag resistance.  The key difference between these two cases is that, in the presence of a magnetic field, the plasmon spectrum becomes gapped at the cyclotron frequency, leading to the suppression of the plasmon contribution to drag resistance. In contrast, in systems with spontaneously broken time-reversal symmetry, we find that plasmons hybridize with odd viscosity waves, and the spectrum remains gapless, as described by Eq. \eqref{eq:mode}.

In Refs. \cite{Patel:2017,Zverevich:2023}, hydrodynamic drag was studied in systems lacking Galilean invariance. Our results generalize this analysis to the case of odd electron liquids. In particular, in the limit $\eta_o \to 0$, our Eq. \eqref{eq:drag-NGI} reproduces the main findings of Ref. \cite{Zverevich:2023}. The question of the crossover from the well-known $T^2$ behavior of drag resistance at low temperatures to the higher-temperature hydrodynamic regime considered here is nontrivial.
This crossover was investigated in detail in Ref. \cite{Chen:2015}, though without accounting for the effects of odd viscosity.

The central results of this work are given by Eqs. \eqref{eq:drag-GI} and \eqref{eq:drag-NGI}, which describe drag resistivity in odd electron liquids with and without Galilean invariance, respectively. We conclude that the effect of odd viscosity in the Galilean-invariant case is weak. In this case, the temperature dependence of drag resistivity is primarily determined by shear viscosity, which governs plasmon attenuation. However, the effect of odd viscosity is stronger in liquids without Galilean invariance. Here, the temperature dependence is dictated by a factor of $\sim T$, arising from thermal fluctuations, and by the intrinsic conductivity $\sigma(T)$. For example, in monolayer graphene, it exhibits only a logarithmic dependence, $\sigma(T) \sim e^2 \ln^2 T$, thus it varies weakly with temperature. Consequently, the temperature dependence of the ratio $\rho_{\text{D}}/T$ is primarily governed by $\eta_o(T)$. Additionally, drag is significantly enhanced in this case numerically and exhibits a slower decay versus the interlayer separation. These features, at least in principle, open the possibility of extracting odd viscosity from Coulomb drag measurements.
The obtained results are also applicable to non-Fermi liquids, as we made no specific assumptions about the temperature dependence of the fluid’s dissipative properties. In this sense, the derived expressions for the drag resistance are generic within the domain of applicability of hydrodynamic theory. We close by noting that additional contributions to the drag force may arise from interlayer skew scattering mediated by Coulomb interactions. This mechanism may lead to a Hall-like response \cite{Messica:2024,Badalyan:2009}, the magnitude of which in the hydrodynamic regime of odd electron liquids remains to be understood.

\section*{Acknowledgments}

This work was supported by the National Science Foundation Grant No. DMR-2452658 and H. I. Romnes Faculty Fellowship provided by the University of Wisconsin-Madison Office of the Vice Chancellor for Research and Graduate Education with funding from the Wisconsin Alumni Research Foundation.

\section*{Data availability} 

The data presented in Figs. 1--3 were generated from analytical expressions derived and defined in the paper. The Mathematica code used to produce these plots will be made available by the authors upon reasonable request.

\bibliography{biblio}

\begin{thebibliography}{40}%
\makeatletter
\providecommand \@ifxundefined [1]{%
 \@ifx{#1\undefined}
}%
\providecommand \@ifnum [1]{%
 \ifnum #1\expandafter \@firstoftwo
 \else \expandafter \@secondoftwo
 \fi
}%
\providecommand \@ifx [1]{%
 \ifx #1\expandafter \@firstoftwo
 \else \expandafter \@secondoftwo
 \fi
}%
\providecommand \natexlab [1]{#1}%
\providecommand \enquote  [1]{``#1''}%
\providecommand \bibnamefont  [1]{#1}%
\providecommand \bibfnamefont [1]{#1}%
\providecommand \citenamefont [1]{#1}%
\providecommand \href@noop [0]{\@secondoftwo}%
\providecommand \href [0]{\begingroup \@sanitize@url \@href}%
\providecommand \@href[1]{\@@startlink{#1}\@@href}%
\providecommand \@@href[1]{\endgroup#1\@@endlink}%
\providecommand \@sanitize@url [0]{\catcode `\\12\catcode `\$12\catcode
  `\&12\catcode `\#12\catcode `\^12\catcode `\_12\catcode `\%12\relax}%
\providecommand \@@startlink[1]{}%
\providecommand \@@endlink[0]{}%
\providecommand \url  [0]{\begingroup\@sanitize@url \@url }%
\providecommand \@url [1]{\endgroup\@href {#1}{\urlprefix }}%
\providecommand \urlprefix  [0]{URL }%
\providecommand \Eprint [0]{\href }%
\providecommand \doibase [0]{http://dx.doi.org/}%
\providecommand \selectlanguage [0]{\@gobble}%
\providecommand \bibinfo  [0]{\@secondoftwo}%
\providecommand \bibfield  [0]{\@secondoftwo}%
\providecommand \translation [1]{[#1]}%
\providecommand \BibitemOpen [0]{}%
\providecommand \bibitemStop [0]{}%
\providecommand \bibitemNoStop [0]{.\EOS\space}%
\providecommand \EOS [0]{\spacefactor3000\relax}%
\providecommand \BibitemShut  [1]{\csname bibitem#1\endcsname}%
\let\auto@bib@innerbib\@empty
%</preamble>
\bibitem [{\citenamefont {Fruchart}\ \emph {et~al.}(2023)\citenamefont
  {Fruchart}, \citenamefont {Scheibner},\ and\ \citenamefont
  {Vitelli}}]{OddReview}%
  \BibitemOpen
  \bibfield  {author} {\bibinfo {author} {\bibfnamefont {Michel}\ \bibnamefont
  {Fruchart}}, \bibinfo {author} {\bibfnamefont {Colin}\ \bibnamefont
  {Scheibner}}, \ and\ \bibinfo {author} {\bibfnamefont {Vincenzo}\
  \bibnamefont {Vitelli}},\ }\bibfield  {title} {\enquote {\bibinfo {title}
  {Odd viscosity and odd elasticity},}\ }\href {\doibase
  https://doi.org/10.1146/annurev-conmatphys-040821-125506} {\bibfield
  {journal} {\bibinfo  {journal} {Annual Review of Condensed Matter Physics}\
  }\textbf {\bibinfo {volume} {14}},\ \bibinfo {pages} {471--510} (\bibinfo
  {year} {2023})}\BibitemShut {NoStop}%
\bibitem [{\citenamefont {Steinberg}(1958)}]{Steinberg:1958}%
  \BibitemOpen
  \bibfield  {author} {\bibinfo {author} {\bibfnamefont {M.~S.}\ \bibnamefont
  {Steinberg}},\ }\bibfield  {title} {\enquote {\bibinfo {title} {Viscosity of
  the electron gas in metals},}\ }\href {\doibase 10.1103/PhysRev.109.1486}
  {\bibfield  {journal} {\bibinfo  {journal} {Phys. Rev.}\ }\textbf {\bibinfo
  {volume} {109}},\ \bibinfo {pages} {1486--1492} (\bibinfo {year}
  {1958})}\BibitemShut {NoStop}%
\bibitem [{\citenamefont {Kaufman}(1960)}]{Kaufman:1960}%
  \BibitemOpen
  \bibfield  {author} {\bibinfo {author} {\bibfnamefont {Allan~N.}\
  \bibnamefont {Kaufman}},\ }\bibfield  {title} {\enquote {\bibinfo {title}
  {Plasma viscosity in a magnetic field},}\ }\href {\doibase 10.1063/1.1706096}
  {\bibfield  {journal} {\bibinfo  {journal} {The Physics of Fluids}\ }\textbf
  {\bibinfo {volume} {3}},\ \bibinfo {pages} {610--616} (\bibinfo {year}
  {1960})}\BibitemShut {NoStop}%
\bibitem [{\citenamefont {Avron}\ \emph {et~al.}(1995)\citenamefont {Avron},
  \citenamefont {Seiler},\ and\ \citenamefont {Zograf}}]{Avron:1995}%
  \BibitemOpen
  \bibfield  {author} {\bibinfo {author} {\bibfnamefont {J.~E.}\ \bibnamefont
  {Avron}}, \bibinfo {author} {\bibfnamefont {R.}~\bibnamefont {Seiler}}, \
  and\ \bibinfo {author} {\bibfnamefont {P.~G.}\ \bibnamefont {Zograf}},\
  }\bibfield  {title} {\enquote {\bibinfo {title} {Viscosity of quantum {H}all
  fluids},}\ }\href {\doibase 10.1103/PhysRevLett.75.697} {\bibfield  {journal}
  {\bibinfo  {journal} {Phys. Rev. Lett.}\ }\textbf {\bibinfo {volume} {75}},\
  \bibinfo {pages} {697--700} (\bibinfo {year} {1995})}\BibitemShut {NoStop}%
\bibitem [{\citenamefont {Avron}(1998)}]{Avron:1998}%
  \BibitemOpen
  \bibfield  {author} {\bibinfo {author} {\bibfnamefont {J.~E.}\ \bibnamefont
  {Avron}},\ }\bibfield  {title} {\enquote {\bibinfo {title} {Odd viscosity},}\
  }\href {\doibase 10.1023/a:1023084404080} {\bibfield  {journal} {\bibinfo
  {journal} {J. Stat. Phys.}\ }\textbf {\bibinfo {volume} {92}},\ \bibinfo
  {pages} {543--557} (\bibinfo {year} {1998})}\BibitemShut {NoStop}%
\bibitem [{\citenamefont {Lucas}\ and\ \citenamefont
  {Fong}(2018)}]{Lucas:2018}%
  \BibitemOpen
  \bibfield  {author} {\bibinfo {author} {\bibfnamefont {Andrew}\ \bibnamefont
  {Lucas}}\ and\ \bibinfo {author} {\bibfnamefont {Kin~Chung}\ \bibnamefont
  {Fong}},\ }\bibfield  {title} {\enquote {\bibinfo {title} {Hydrodynamics of
  electrons in graphene},}\ }\href {\doibase 10.1088/1361-648X/aaa274}
  {\bibfield  {journal} {\bibinfo  {journal} {Journal of Physics: Condensed
  Matter}\ }\textbf {\bibinfo {volume} {30}},\ \bibinfo {pages} {053001}
  (\bibinfo {year} {2018})}\BibitemShut {NoStop}%
\bibitem [{\citenamefont {Levchenko}\ and\ \citenamefont
  {Schmalian}(2020)}]{Levchenko:2020}%
  \BibitemOpen
  \bibfield  {author} {\bibinfo {author} {\bibfnamefont {Alex}\ \bibnamefont
  {Levchenko}}\ and\ \bibinfo {author} {\bibfnamefont {J{\"o}rg}\ \bibnamefont
  {Schmalian}},\ }\bibfield  {title} {\enquote {\bibinfo {title} {Transport
  properties of strongly coupled electron--phonon liquids},}\ }\href {\doibase
  https://doi.org/10.1016/j.aop.2020.168218} {\bibfield  {journal} {\bibinfo
  {journal} {Annals of Physics}\ }\textbf {\bibinfo {volume} {419}},\ \bibinfo
  {pages} {168218} (\bibinfo {year} {2020})}\BibitemShut {NoStop}%
\bibitem [{\citenamefont {Narozhny}(2022)}]{Narozhny:2022}%
  \BibitemOpen
  \bibfield  {author} {\bibinfo {author} {\bibfnamefont {Boris~N.}\
  \bibnamefont {Narozhny}},\ }\bibfield  {title} {\enquote {\bibinfo {title}
  {Hydrodynamic approach to two-dimensional electron systems},}\ }\href
  {\doibase 10.1007/s40766-022-00036-z} {\bibfield  {journal} {\bibinfo
  {journal} {La Rivista del Nuovo Cimento}\ }\textbf {\bibinfo {volume} {45}},\
  \bibinfo {pages} {661--736} (\bibinfo {year} {2022})}\BibitemShut {NoStop}%
\bibitem [{\citenamefont {Fritz}\ and\ \citenamefont
  {Scaffidi}(2024)}]{Fritz:2024}%
  \BibitemOpen
  \bibfield  {author} {\bibinfo {author} {\bibfnamefont {L.}~\bibnamefont
  {Fritz}}\ and\ \bibinfo {author} {\bibfnamefont {T.}~\bibnamefont
  {Scaffidi}},\ }\bibfield  {title} {\enquote {\bibinfo {title} {Hydrodynamic
  electronic transport},}\ }\href {\doibase
  https://doi.org/10.1146/annurev-conmatphys-040521-042014} {\bibfield
  {journal} {\bibinfo  {journal} {Annual Review of Condensed Matter Physics}\
  }\textbf {\bibinfo {volume} {15}},\ \bibinfo {pages} {17--44} (\bibinfo
  {year} {2024})}\BibitemShut {NoStop}%
\bibitem [{\citenamefont {Hoyos}\ and\ \citenamefont {Son}(2012)}]{Hoyos:2012}%
  \BibitemOpen
  \bibfield  {author} {\bibinfo {author} {\bibfnamefont {Carlos}\ \bibnamefont
  {Hoyos}}\ and\ \bibinfo {author} {\bibfnamefont {Dam~Thanh}\ \bibnamefont
  {Son}},\ }\bibfield  {title} {\enquote {\bibinfo {title} {Hall viscosity and
  electromagnetic response},}\ }\href {\doibase 10.1103/PhysRevLett.108.066805}
  {\bibfield  {journal} {\bibinfo  {journal} {Phys. Rev. Lett.}\ }\textbf
  {\bibinfo {volume} {108}},\ \bibinfo {pages} {066805} (\bibinfo {year}
  {2012})}\BibitemShut {NoStop}%
\bibitem [{\citenamefont {Sherafati}\ \emph {et~al.}(2016)\citenamefont
  {Sherafati}, \citenamefont {Principi},\ and\ \citenamefont
  {Vignale}}]{Vignale:2016}%
  \BibitemOpen
  \bibfield  {author} {\bibinfo {author} {\bibfnamefont {Mohammad}\
  \bibnamefont {Sherafati}}, \bibinfo {author} {\bibfnamefont {Alessandro}\
  \bibnamefont {Principi}}, \ and\ \bibinfo {author} {\bibfnamefont {Giovanni}\
  \bibnamefont {Vignale}},\ }\bibfield  {title} {\enquote {\bibinfo {title}
  {Hall viscosity and electromagnetic response of electrons in graphene},}\
  }\href {\doibase 10.1103/PhysRevB.94.125427} {\bibfield  {journal} {\bibinfo
  {journal} {Phys. Rev. B}\ }\textbf {\bibinfo {volume} {94}},\ \bibinfo
  {pages} {125427} (\bibinfo {year} {2016})}\BibitemShut {NoStop}%
\bibitem [{\citenamefont {Scaffidi}\ \emph {et~al.}(2017)\citenamefont
  {Scaffidi}, \citenamefont {Nandi}, \citenamefont {Schmidt}, \citenamefont
  {Mackenzie},\ and\ \citenamefont {Moore}}]{Scaffidi:2017}%
  \BibitemOpen
  \bibfield  {author} {\bibinfo {author} {\bibfnamefont {Thomas}\ \bibnamefont
  {Scaffidi}}, \bibinfo {author} {\bibfnamefont {Nabhanila}\ \bibnamefont
  {Nandi}}, \bibinfo {author} {\bibfnamefont {Burkhard}\ \bibnamefont
  {Schmidt}}, \bibinfo {author} {\bibfnamefont {Andrew~P.}\ \bibnamefont
  {Mackenzie}}, \ and\ \bibinfo {author} {\bibfnamefont {Joel~E.}\ \bibnamefont
  {Moore}},\ }\bibfield  {title} {\enquote {\bibinfo {title} {Hydrodynamic
  electron flow and hall viscosity},}\ }\href {\doibase
  10.1103/PhysRevLett.118.226601} {\bibfield  {journal} {\bibinfo  {journal}
  {Phys. Rev. Lett.}\ }\textbf {\bibinfo {volume} {118}},\ \bibinfo {pages}
  {226601} (\bibinfo {year} {2017})}\BibitemShut {NoStop}%
\bibitem [{\citenamefont {Delacr\'etaz}\ and\ \citenamefont
  {Gromov}(2017)}]{Delacretaz:2017}%
  \BibitemOpen
  \bibfield  {author} {\bibinfo {author} {\bibfnamefont {Luca~V.}\ \bibnamefont
  {Delacr\'etaz}}\ and\ \bibinfo {author} {\bibfnamefont {Andrey}\ \bibnamefont
  {Gromov}},\ }\bibfield  {title} {\enquote {\bibinfo {title} {Transport
  signatures of the {H}all viscosity},}\ }\href {\doibase
  10.1103/PhysRevLett.119.226602} {\bibfield  {journal} {\bibinfo  {journal}
  {Phys. Rev. Lett.}\ }\textbf {\bibinfo {volume} {119}},\ \bibinfo {pages}
  {226602} (\bibinfo {year} {2017})}\BibitemShut {NoStop}%
\bibitem [{\citenamefont {Pellegrino}\ \emph {et~al.}(2017)\citenamefont
  {Pellegrino}, \citenamefont {Torre},\ and\ \citenamefont
  {Polini}}]{Polini:2017}%
  \BibitemOpen
  \bibfield  {author} {\bibinfo {author} {\bibfnamefont {Francesco M.~D.}\
  \bibnamefont {Pellegrino}}, \bibinfo {author} {\bibfnamefont {Iacopo}\
  \bibnamefont {Torre}}, \ and\ \bibinfo {author} {\bibfnamefont {Marco}\
  \bibnamefont {Polini}},\ }\bibfield  {title} {\enquote {\bibinfo {title}
  {Nonlocal transport and the {H}all viscosity of two-dimensional hydrodynamic
  electron liquids},}\ }\href {\doibase 10.1103/PhysRevB.96.195401} {\bibfield
  {journal} {\bibinfo  {journal} {Phys. Rev. B}\ }\textbf {\bibinfo {volume}
  {96}},\ \bibinfo {pages} {195401} (\bibinfo {year} {2017})}\BibitemShut
  {NoStop}%
\bibitem [{\citenamefont {Holder}\ \emph {et~al.}(2019)\citenamefont {Holder},
  \citenamefont {Queiroz},\ and\ \citenamefont {Stern}}]{Holder:2019}%
  \BibitemOpen
  \bibfield  {author} {\bibinfo {author} {\bibfnamefont {Tobias}\ \bibnamefont
  {Holder}}, \bibinfo {author} {\bibfnamefont {Raquel}\ \bibnamefont
  {Queiroz}}, \ and\ \bibinfo {author} {\bibfnamefont {Ady}\ \bibnamefont
  {Stern}},\ }\bibfield  {title} {\enquote {\bibinfo {title} {Unified
  description of the classical {H}all viscosity},}\ }\href {\doibase
  10.1103/PhysRevLett.123.106801} {\bibfield  {journal} {\bibinfo  {journal}
  {Phys. Rev. Lett.}\ }\textbf {\bibinfo {volume} {123}},\ \bibinfo {pages}
  {106801} (\bibinfo {year} {2019})}\BibitemShut {NoStop}%
\bibitem [{\citenamefont {Berdyugin}\ \emph {et~al.}(2019)\citenamefont
  {Berdyugin}, \citenamefont {Xu}, \citenamefont {Pellegrino}, \citenamefont
  {Kumar}, \citenamefont {Principi}, \citenamefont {Torre}, \citenamefont
  {Shalom}, \citenamefont {Taniguchi}, \citenamefont {Watanabe}, \citenamefont
  {Grigorieva}, \citenamefont {Polini}, \citenamefont {Geim},\ and\
  \citenamefont {Bandurin}}]{Berdyugin:2019}%
  \BibitemOpen
  \bibfield  {author} {\bibinfo {author} {\bibfnamefont {A.~I.}\ \bibnamefont
  {Berdyugin}}, \bibinfo {author} {\bibfnamefont {S.~G.}\ \bibnamefont {Xu}},
  \bibinfo {author} {\bibfnamefont {F.~M.~D.}\ \bibnamefont {Pellegrino}},
  \bibinfo {author} {\bibfnamefont {R.~Krishna}\ \bibnamefont {Kumar}},
  \bibinfo {author} {\bibfnamefont {A.}~\bibnamefont {Principi}}, \bibinfo
  {author} {\bibfnamefont {I.}~\bibnamefont {Torre}}, \bibinfo {author}
  {\bibfnamefont {M.~Ben}\ \bibnamefont {Shalom}}, \bibinfo {author}
  {\bibfnamefont {T.}~\bibnamefont {Taniguchi}}, \bibinfo {author}
  {\bibfnamefont {K.}~\bibnamefont {Watanabe}}, \bibinfo {author}
  {\bibfnamefont {I.~V.}\ \bibnamefont {Grigorieva}}, \bibinfo {author}
  {\bibfnamefont {M.}~\bibnamefont {Polini}}, \bibinfo {author} {\bibfnamefont
  {A.~K.}\ \bibnamefont {Geim}}, \ and\ \bibinfo {author} {\bibfnamefont
  {D.~A.}\ \bibnamefont {Bandurin}},\ }\bibfield  {title} {\enquote {\bibinfo
  {title} {Measuring {H}all viscosity of graphene's electron fluid},}\ }\href
  {\doibase 10.1126/science.aau0685} {\bibfield  {journal} {\bibinfo  {journal}
  {Science}\ }\textbf {\bibinfo {volume} {364}},\ \bibinfo {pages} {162--165}
  (\bibinfo {year} {2019})}\BibitemShut {NoStop}%
\bibitem [{\citenamefont {Zhou}\ \emph {et~al.}(2021)\citenamefont {Zhou},
  \citenamefont {Xie}, \citenamefont {Ghazaryan}, \citenamefont {Holder},
  \citenamefont {Ehrets}, \citenamefont {Spanton}, \citenamefont {Taniguchi},
  \citenamefont {Watanabe}, \citenamefont {Berg}, \citenamefont {Serbyn},\ and\
  \citenamefont {Young}}]{Zhou:2021}%
  \BibitemOpen
  \bibfield  {author} {\bibinfo {author} {\bibfnamefont {Haoxin}\ \bibnamefont
  {Zhou}}, \bibinfo {author} {\bibfnamefont {Tian}\ \bibnamefont {Xie}},
  \bibinfo {author} {\bibfnamefont {Areg}\ \bibnamefont {Ghazaryan}}, \bibinfo
  {author} {\bibfnamefont {Tobias}\ \bibnamefont {Holder}}, \bibinfo {author}
  {\bibfnamefont {James~R.}\ \bibnamefont {Ehrets}}, \bibinfo {author}
  {\bibfnamefont {Eric~M.}\ \bibnamefont {Spanton}}, \bibinfo {author}
  {\bibfnamefont {Takashi}\ \bibnamefont {Taniguchi}}, \bibinfo {author}
  {\bibfnamefont {Kenji}\ \bibnamefont {Watanabe}}, \bibinfo {author}
  {\bibfnamefont {Erez}\ \bibnamefont {Berg}}, \bibinfo {author} {\bibfnamefont
  {Maksym}\ \bibnamefont {Serbyn}}, \ and\ \bibinfo {author} {\bibfnamefont
  {Andrea~F.}\ \bibnamefont {Young}},\ }\bibfield  {title} {\enquote {\bibinfo
  {title} {Half- and quarter-metals in rhombohedral trilayer graphene},}\
  }\href {\doibase 10.1038/s41586-021-03938-w} {\bibfield  {journal} {\bibinfo
  {journal} {Nature}\ }\textbf {\bibinfo {volume} {598}},\ \bibinfo {pages}
  {429--433} (\bibinfo {year} {2021})}\BibitemShut {NoStop}%
\bibitem [{\citenamefont {Narozhny}\ and\ \citenamefont
  {Levchenko}(2016)}]{Review:2016}%
  \BibitemOpen
  \bibfield  {author} {\bibinfo {author} {\bibfnamefont {B.~N.}\ \bibnamefont
  {Narozhny}}\ and\ \bibinfo {author} {\bibfnamefont {A.}~\bibnamefont
  {Levchenko}},\ }\bibfield  {title} {\enquote {\bibinfo {title} {Coulomb
  drag},}\ }\href {\doibase 10.1103/RevModPhys.88.025003} {\bibfield  {journal}
  {\bibinfo  {journal} {Rev. Mod. Phys.}\ }\textbf {\bibinfo {volume} {88}},\
  \bibinfo {pages} {025003} (\bibinfo {year} {2016})}\BibitemShut {NoStop}%
\bibitem [{\citenamefont {Apostolov}\ \emph {et~al.}(2014)\citenamefont
  {Apostolov}, \citenamefont {Levchenko},\ and\ \citenamefont
  {Andreev}}]{Apostolov:2014}%
  \BibitemOpen
  \bibfield  {author} {\bibinfo {author} {\bibfnamefont {S.~S.}\ \bibnamefont
  {Apostolov}}, \bibinfo {author} {\bibfnamefont {A.}~\bibnamefont
  {Levchenko}}, \ and\ \bibinfo {author} {\bibfnamefont {A.~V.}\ \bibnamefont
  {Andreev}},\ }\bibfield  {title} {\enquote {\bibinfo {title} {Hydrodynamic
  {C}oulomb drag of strongly correlated electron liquids},}\ }\href {\doibase
  10.1103/PhysRevB.89.121104} {\bibfield  {journal} {\bibinfo  {journal} {Phys.
  Rev. B}\ }\textbf {\bibinfo {volume} {89}},\ \bibinfo {pages} {121104}
  (\bibinfo {year} {2014})}\BibitemShut {NoStop}%
\bibitem [{\citenamefont {Landau}\ and\ \citenamefont
  {Lifshitz}(1957)}]{LL:1957}%
  \BibitemOpen
  \bibfield  {author} {\bibinfo {author} {\bibfnamefont {L.~D.}\ \bibnamefont
  {Landau}}\ and\ \bibinfo {author} {\bibfnamefont {E.~M.}\ \bibnamefont
  {Lifshitz}},\ }\bibfield  {title} {\enquote {\bibinfo {title} {Hydrodynamic
  fluctuations},}\ }\href@noop {} {\bibfield  {journal} {\bibinfo  {journal}
  {JETP}\ }\textbf {\bibinfo {volume} {5}},\ \bibinfo {pages} {512} (\bibinfo
  {year} {1957})}\BibitemShut {NoStop}%
\bibitem [{\citenamefont {Khalatnikov}(1958)}]{Khalatnikov:1958}%
  \BibitemOpen
  \bibfield  {author} {\bibinfo {author} {\bibfnamefont {I.~M.}\ \bibnamefont
  {Khalatnikov}},\ }\bibfield  {title} {\enquote {\bibinfo {title}
  {Hydrodynamic fluctuations in a superfluid liquid},}\ }\href@noop {}
  {\bibfield  {journal} {\bibinfo  {journal} {JETP}\ }\textbf {\bibinfo
  {volume} {6}},\ \bibinfo {pages} {624} (\bibinfo {year} {1958})}\BibitemShut
  {NoStop}%
\bibitem [{\citenamefont {Forster}(1975)}]{Forster}%
  \BibitemOpen
  \bibfield  {author} {\bibinfo {author} {\bibfnamefont {Dieter}\ \bibnamefont
  {Forster}},\ }\href@noop {} {\emph {\bibinfo {title} {Hydrodynamic
  Fluctuations, Broken Symmetry, and Correlation Functions}}}\ (\bibinfo
  {publisher} {W. A. Benjamin, Advanced Book Program},\ \bibinfo {year}
  {1975})\BibitemShut {NoStop}%
\bibitem [{\citenamefont {Lifshitz}\ and\ \citenamefont
  {Pitaevskii}(1980)}]{LL-V9}%
  \BibitemOpen
  \bibfield  {author} {\bibinfo {author} {\bibfnamefont {E.~M.}\ \bibnamefont
  {Lifshitz}}\ and\ \bibinfo {author} {\bibfnamefont {L.~P.}\ \bibnamefont
  {Pitaevskii}},\ }\href@noop {} {\emph {\bibinfo {title} {Statistical Physics,
  Part 2}}},\ \bibinfo {edition} {3rd}\ ed.,\ \bibinfo {series} {Course of
  Theoretical Physics}, Vol.~\bibinfo {volume} {9}\ (\bibinfo  {publisher}
  {Pergamon Press},\ \bibinfo {year} {1980})\BibitemShut {NoStop}%
\bibitem [{\citenamefont {Kogan}(1996)}]{Kogan:1996}%
  \BibitemOpen
  \bibfield  {author} {\bibinfo {author} {\bibfnamefont {Sh.}\ \bibnamefont
  {Kogan}},\ }\href@noop {} {\emph {\bibinfo {title} {Electronic Noise and
  Fluctuations in Solids}}},\ \bibinfo {edition} {1st}\ ed.\ (\bibinfo
  {publisher} {Cambridge University Press},\ \bibinfo {year}
  {1996})\BibitemShut {NoStop}%
\bibitem [{\citenamefont {Chen}\ \emph {et~al.}(2015)\citenamefont {Chen},
  \citenamefont {Andreev},\ and\ \citenamefont {Levchenko}}]{Chen:2015}%
  \BibitemOpen
  \bibfield  {author} {\bibinfo {author} {\bibfnamefont {W.}~\bibnamefont
  {Chen}}, \bibinfo {author} {\bibfnamefont {A.~V.}\ \bibnamefont {Andreev}}, \
  and\ \bibinfo {author} {\bibfnamefont {A.}~\bibnamefont {Levchenko}},\
  }\bibfield  {title} {\enquote {\bibinfo {title} {Boltzmann-{L}angevin theory
  of {C}oulomb drag},}\ }\href {\doibase 10.1103/PhysRevB.91.245405} {\bibfield
   {journal} {\bibinfo  {journal} {Phys. Rev. B}\ }\textbf {\bibinfo {volume}
  {91}},\ \bibinfo {pages} {245405} (\bibinfo {year} {2015})}\BibitemShut
  {NoStop}%
\bibitem [{\citenamefont {Landau}\ and\ \citenamefont
  {Lifshitz}(1987)}]{LL-V6}%
  \BibitemOpen
  \bibfield  {author} {\bibinfo {author} {\bibfnamefont {L.~D.}\ \bibnamefont
  {Landau}}\ and\ \bibinfo {author} {\bibfnamefont {E.~M.}\ \bibnamefont
  {Lifshitz}},\ }\href@noop {} {\emph {\bibinfo {title} {Fluid Mechanics}}},\
  \bibinfo {edition} {2nd}\ ed.,\ \bibinfo {series} {Course of Theoretical
  Physics Series}, Vol.~\bibinfo {volume} {6}\ (\bibinfo  {publisher}
  {Butterworth-Heinemann, Oxford},\ \bibinfo {year} {1987})\BibitemShut
  {NoStop}%
\bibitem [{\citenamefont {Fritz}\ \emph {et~al.}(2008)\citenamefont {Fritz},
  \citenamefont {Schmalian}, \citenamefont {M\"uller},\ and\ \citenamefont
  {Sachdev}}]{Fritz:2008}%
  \BibitemOpen
  \bibfield  {author} {\bibinfo {author} {\bibfnamefont {Lars}\ \bibnamefont
  {Fritz}}, \bibinfo {author} {\bibfnamefont {J\"org}\ \bibnamefont
  {Schmalian}}, \bibinfo {author} {\bibfnamefont {Markus}\ \bibnamefont
  {M\"uller}}, \ and\ \bibinfo {author} {\bibfnamefont {Subir}\ \bibnamefont
  {Sachdev}},\ }\bibfield  {title} {\enquote {\bibinfo {title} {Quantum
  critical transport in clean graphene},}\ }\href {\doibase
  10.1103/PhysRevB.78.085416} {\bibfield  {journal} {\bibinfo  {journal} {Phys.
  Rev. B}\ }\textbf {\bibinfo {volume} {78}},\ \bibinfo {pages} {085416}
  (\bibinfo {year} {2008})}\BibitemShut {NoStop}%
\bibitem [{\citenamefont {Kashuba}(2008)}]{Kashuba:2008}%
  \BibitemOpen
  \bibfield  {author} {\bibinfo {author} {\bibfnamefont {Alexander~B.}\
  \bibnamefont {Kashuba}},\ }\bibfield  {title} {\enquote {\bibinfo {title}
  {Conductivity of defectless graphene},}\ }\href {\doibase
  10.1103/PhysRevB.78.085415} {\bibfield  {journal} {\bibinfo  {journal} {Phys.
  Rev. B}\ }\textbf {\bibinfo {volume} {78}},\ \bibinfo {pages} {085415}
  (\bibinfo {year} {2008})}\BibitemShut {NoStop}%
\bibitem [{\citenamefont {Ganeshan}\ and\ \citenamefont
  {Abanov}(2017)}]{Ganeshan:2017}%
  \BibitemOpen
  \bibfield  {author} {\bibinfo {author} {\bibfnamefont {Sriram}\ \bibnamefont
  {Ganeshan}}\ and\ \bibinfo {author} {\bibfnamefont {Alexander~G.}\
  \bibnamefont {Abanov}},\ }\bibfield  {title} {\enquote {\bibinfo {title} {Odd
  viscosity in two-dimensional incompressible fluids},}\ }\href {\doibase
  10.1103/PhysRevFluids.2.094101} {\bibfield  {journal} {\bibinfo  {journal}
  {Phys. Rev. Fluids}\ }\textbf {\bibinfo {volume} {2}},\ \bibinfo {pages}
  {094101} (\bibinfo {year} {2017})}\BibitemShut {NoStop}%
\bibitem [{\citenamefont {Lifshitz}\ and\ \citenamefont
  {Pitaevskii}(1981)}]{LL-V10}%
  \BibitemOpen
  \bibfield  {author} {\bibinfo {author} {\bibfnamefont {E.~M.}\ \bibnamefont
  {Lifshitz}}\ and\ \bibinfo {author} {\bibfnamefont {L.~P.}\ \bibnamefont
  {Pitaevskii}},\ }\href@noop {} {\emph {\bibinfo {title} {Physical
  Kinetics}}},\ \bibinfo {edition} {1st}\ ed.,\ \bibinfo {series} {Course of
  Theoretical Physics Series}, Vol.~\bibinfo {volume} {10}\ (\bibinfo
  {publisher} {Butterworth-Heinemann},\ \bibinfo {year} {1981})\BibitemShut
  {NoStop}%
\bibitem [{\citenamefont {Hruska}\ and\ \citenamefont
  {Spivak}(2002)}]{Hruska:2002}%
  \BibitemOpen
  \bibfield  {author} {\bibinfo {author} {\bibfnamefont {M.}~\bibnamefont
  {Hruska}}\ and\ \bibinfo {author} {\bibfnamefont {B.}~\bibnamefont
  {Spivak}},\ }\bibfield  {title} {\enquote {\bibinfo {title} {Conductivity of
  the classical two-dimensional electron gas},}\ }\href {\doibase
  10.1103/PhysRevB.65.033315} {\bibfield  {journal} {\bibinfo  {journal} {Phys.
  Rev. B}\ }\textbf {\bibinfo {volume} {65}},\ \bibinfo {pages} {033315}
  (\bibinfo {year} {2002})}\BibitemShut {NoStop}%
\bibitem [{\citenamefont {Patel}\ \emph {et~al.}(2017)\citenamefont {Patel},
  \citenamefont {Davison},\ and\ \citenamefont {Levchenko}}]{Patel:2017}%
  \BibitemOpen
  \bibfield  {author} {\bibinfo {author} {\bibfnamefont {Aavishkar~A.}\
  \bibnamefont {Patel}}, \bibinfo {author} {\bibfnamefont {Richard~A.}\
  \bibnamefont {Davison}}, \ and\ \bibinfo {author} {\bibfnamefont {Alex}\
  \bibnamefont {Levchenko}},\ }\bibfield  {title} {\enquote {\bibinfo {title}
  {Hydrodynamic flows of non-{F}ermi liquids: Magnetotransport and bilayer
  drag},}\ }\href {\doibase 10.1103/PhysRevB.96.205417} {\bibfield  {journal}
  {\bibinfo  {journal} {Phys. Rev. B}\ }\textbf {\bibinfo {volume} {96}},\
  \bibinfo {pages} {205417} (\bibinfo {year} {2017})}\BibitemShut {NoStop}%
\bibitem [{\citenamefont {Sinitsyn}\ \emph {et~al.}(2007)\citenamefont
  {Sinitsyn}, \citenamefont {MacDonald}, \citenamefont {Jungwirth},
  \citenamefont {Dugaev},\ and\ \citenamefont {Sinova}}]{Sinitsyn:2007}%
  \BibitemOpen
  \bibfield  {author} {\bibinfo {author} {\bibfnamefont {N.~A.}\ \bibnamefont
  {Sinitsyn}}, \bibinfo {author} {\bibfnamefont {A.~H.}\ \bibnamefont
  {MacDonald}}, \bibinfo {author} {\bibfnamefont {T.}~\bibnamefont
  {Jungwirth}}, \bibinfo {author} {\bibfnamefont {V.~K.}\ \bibnamefont
  {Dugaev}}, \ and\ \bibinfo {author} {\bibfnamefont {Jairo}\ \bibnamefont
  {Sinova}},\ }\bibfield  {title} {\enquote {\bibinfo {title} {Anomalous {H}all
  effect in a two-dimensional {D}irac band: The link between the
  {K}ubo-{S}treda formula and the semiclassical {B}oltzmann equation
  approach},}\ }\href {\doibase 10.1103/PhysRevB.75.045315} {\bibfield
  {journal} {\bibinfo  {journal} {Phys. Rev. B}\ }\textbf {\bibinfo {volume}
  {75}},\ \bibinfo {pages} {045315} (\bibinfo {year} {2007})}\BibitemShut
  {NoStop}%
\bibitem [{\citenamefont {K{\"o}nig}\ and\ \citenamefont
  {Levchenko}(2021)}]{Konig:2021}%
  \BibitemOpen
  \bibfield  {author} {\bibinfo {author} {\bibfnamefont {Elio~J.}\ \bibnamefont
  {K{\"o}nig}}\ and\ \bibinfo {author} {\bibfnamefont {Alex}\ \bibnamefont
  {Levchenko}},\ }\bibfield  {title} {\enquote {\bibinfo {title} {Quantum
  kinetics of anomalous and nonlinear {H}all effects in topological
  semimetals},}\ }\href {\doibase https://doi.org/10.1016/j.aop.2021.168492}
  {\bibfield  {journal} {\bibinfo  {journal} {Annals of Physics}\ }\textbf
  {\bibinfo {volume} {435}},\ \bibinfo {pages} {168492} (\bibinfo {year}
  {2021})},\ \bibinfo {note} {special issue on Philip W. Anderson}\BibitemShut
  {NoStop}%
\bibitem [{\citenamefont {Glazov}\ and\ \citenamefont
  {Golub}(2022)}]{Golub:2022}%
  \BibitemOpen
  \bibfield  {author} {\bibinfo {author} {\bibfnamefont {M.~M.}\ \bibnamefont
  {Glazov}}\ and\ \bibinfo {author} {\bibfnamefont {L.~E.}\ \bibnamefont
  {Golub}},\ }\bibfield  {title} {\enquote {\bibinfo {title} {Spin and valley
  {H}all effects induced by asymmetric interparticle scattering},}\ }\href
  {\doibase 10.1103/PhysRevB.106.235305} {\bibfield  {journal} {\bibinfo
  {journal} {Phys. Rev. B}\ }\textbf {\bibinfo {volume} {106}},\ \bibinfo
  {pages} {235305} (\bibinfo {year} {2022})}\BibitemShut {NoStop}%
\bibitem [{\citenamefont {Messica}\ and\ \citenamefont
  {Gutman}(2024)}]{Messica:2024}%
  \BibitemOpen
  \bibfield  {author} {\bibinfo {author} {\bibfnamefont {Yonatan}\ \bibnamefont
  {Messica}}\ and\ \bibinfo {author} {\bibfnamefont {Dmitri~B.}\ \bibnamefont
  {Gutman}},\ }\bibfield  {title} {\enquote {\bibinfo {title} {Hall {C}oulomb
  drag induced by electron-electron skew scattering},}\ }\href {\doibase
  10.1103/PhysRevB.110.115424} {\bibfield  {journal} {\bibinfo  {journal}
  {Phys. Rev. B}\ }\textbf {\bibinfo {volume} {110}},\ \bibinfo {pages}
  {115424} (\bibinfo {year} {2024})}\BibitemShut {NoStop}%
\bibitem [{\citenamefont {Apostolov}\ \emph {et~al.}(2019)\citenamefont
  {Apostolov}, \citenamefont {Pesin},\ and\ \citenamefont
  {Levchenko}}]{Apostolov:2019}%
  \BibitemOpen
  \bibfield  {author} {\bibinfo {author} {\bibfnamefont {S.~S.}\ \bibnamefont
  {Apostolov}}, \bibinfo {author} {\bibfnamefont {D.~A.}\ \bibnamefont
  {Pesin}}, \ and\ \bibinfo {author} {\bibfnamefont {A.}~\bibnamefont
  {Levchenko}},\ }\bibfield  {title} {\enquote {\bibinfo {title} {Magnetodrag
  in the hydrodynamic regime: Effects of magnetoplasmon resonance and {H}all
  viscosity},}\ }\href {\doibase 10.1103/PhysRevB.100.115401} {\bibfield
  {journal} {\bibinfo  {journal} {Phys. Rev. B}\ }\textbf {\bibinfo {volume}
  {100}},\ \bibinfo {pages} {115401} (\bibinfo {year} {2019})}\BibitemShut
  {NoStop}%
\bibitem [{\citenamefont {Hasdeo}\ \emph {et~al.}(2023)\citenamefont {Hasdeo},
  \citenamefont {Idrisov},\ and\ \citenamefont {Schmidt}}]{Schmidt:2023}%
  \BibitemOpen
  \bibfield  {author} {\bibinfo {author} {\bibfnamefont {Eddwi~H.}\
  \bibnamefont {Hasdeo}}, \bibinfo {author} {\bibfnamefont {Edvin~G.}\
  \bibnamefont {Idrisov}}, \ and\ \bibinfo {author} {\bibfnamefont {Thomas~L.}\
  \bibnamefont {Schmidt}},\ }\bibfield  {title} {\enquote {\bibinfo {title}
  {Coulomb drag of viscous electron fluids: Drag viscosity and negative drag
  conductivity},}\ }\href {\doibase 10.1103/PhysRevB.107.L121107} {\bibfield
  {journal} {\bibinfo  {journal} {Phys. Rev. B}\ }\textbf {\bibinfo {volume}
  {107}},\ \bibinfo {pages} {L121107} (\bibinfo {year} {2023})}\BibitemShut
  {NoStop}%
\bibitem [{\citenamefont {Zverevich}\ and\ \citenamefont
  {Levchenko}(2023)}]{Zverevich:2023}%
  \BibitemOpen
  \bibfield  {author} {\bibinfo {author} {\bibfnamefont {Dmitry}\ \bibnamefont
  {Zverevich}}\ and\ \bibinfo {author} {\bibfnamefont {Alex}\ \bibnamefont
  {Levchenko}},\ }\bibfield  {title} {\enquote {\bibinfo {title} {Transport
  signatures of plasmon fluctuations in electron hydrodynamics},}\ }\href
  {\doibase 10.1063/10.0022363} {\bibfield  {journal} {\bibinfo  {journal} {Low
  Temperature Physics}\ }\textbf {\bibinfo {volume} {49}},\ \bibinfo {pages}
  {1376--1384} (\bibinfo {year} {2023})}\BibitemShut {NoStop}%
\bibitem [{\citenamefont {Badalyan}\ and\ \citenamefont
  {Vignale}(2009)}]{Badalyan:2009}%
  \BibitemOpen
  \bibfield  {author} {\bibinfo {author} {\bibfnamefont {S.~M.}\ \bibnamefont
  {Badalyan}}\ and\ \bibinfo {author} {\bibfnamefont {G.}~\bibnamefont
  {Vignale}},\ }\bibfield  {title} {\enquote {\bibinfo {title} {Spin {H}all
  drag in electronic bilayers},}\ }\href {\doibase
  10.1103/PhysRevLett.103.196601} {\bibfield  {journal} {\bibinfo  {journal}
  {Phys. Rev. Lett.}\ }\textbf {\bibinfo {volume} {103}},\ \bibinfo {pages}
  {196601} (\bibinfo {year} {2009})}\BibitemShut {NoStop}%
\end{thebibliography}%

\end{document}